\documentclass[sn-vancouver,Numbered]{sn-jnl}

\usepackage{graphicx}%
\usepackage{multirow}%
\usepackage{amsmath,amssymb,amsfonts}%
\usepackage{amsthm}%
\usepackage{mathrsfs}%
\usepackage[title]{appendix}%
\usepackage{xcolor}%
\usepackage{textcomp}%
\usepackage{manyfoot}%
\usepackage{booktabs}%
\usepackage{algorithm}%
\usepackage{algorithmicx}%
\usepackage{algpseudocode}%
\usepackage{listings}%

\theoremstyle{thmstyleone}%
\theoremstyle{thmstyletwo}%

\theoremstyle{thmstylethree}%

\begin{document}

\title[Article Title]{Dark Energy Search by Atom Interferometry in the Einstein-Elevator}


\author*[1]{\fnm{Charles} \sur{Garcion}}\email{garcion@iqo.uni-hannover.de}
\author[1]{\fnm{Sukhjovan S.} \sur{Gill}}
\author[1]{\fnm{Magdalena} \sur{Misslisch}}
\author[4]{\fnm{Alexander} \sur{Heidt}}
\author[2]{\fnm{Ioannis} \sur{Papadakis}}
\author[1,6]{\fnm{Baptist} \sur{Piest}}
\author[2]{\fnm{Vladimir} \sur{Schkolnik}}
\author[1]{\fnm{Thijs} \sur{Wendrich}}
\author[7]{\fnm{Arnau} \sur{Prat}}
\author[7]{\fnm{Kai} \sur{Bleeke}}
\author[2,5]{\fnm{Markus} \sur{Krutzik}}
\author[3]{\fnm{Sheng-wey} \sur{Chiow}}
\author[3]{\fnm{Nan} \sur{Yu}}
\author[4]{\fnm{Christoph} \sur{Lotz}}
\author[1]{\fnm{Naceur} \sur{Gaaloul}}
\author[1]{\fnm{Ernst M.} \sur{Rasel}}



\affil*[1]{\orgdiv{Institut für Quantenoptik}, \orgname{Leibniz Universität Hannover}, \orgaddress{\street{Welfengarten 1}, \city{Hannover}, \postcode{30167}, \country{Germany}}}

\affil[2]{\orgdiv{Institut für Physik}, \orgname{Humboldt Universität zu Berlin}, \orgaddress{\street{Newtonstraße 15},  \city{Berlin}, \postcode{12489}, \country{Germany}}}

\affil[3]{\orgdiv{Jet Propulsion Laboratory}, \orgname{California Institute of Technology}, \orgaddress{ \city{Pasadena}, \postcode{91109}, \state{California}, \country{USA}}}

\affil[4]{\orgdiv{Institut für Transport‑ und Automatisierungstechnik c/o Hannover Institute of Technology}, \orgname{Leibniz Universität Hannover}, \orgaddress{\street{Callinstraße 36}, \city{Hannover}, \postcode{30167}, \country{Germany}}}

\affil[5]{\orgdiv{Ferdinand-Braun-Institut}, \orgaddress{\street{Gustav-Kirchoff-Str. 4},  \city{Berlin}, \postcode{12489}, \country{Germany}}}

\affil[6]{\orgdiv{SYRTE, Observatoire de Paris}, \orgname{Université PSL, CNRS, 
Sorbonne Université}, \orgaddress{\street{61 Avenue de
l’Observatoire}, \city{Paris}, \postcode{75014}, \country{France}}}

\affil[7]{\orgdiv{Institut für Softwaretechnologie}, \orgname{Deutsches Zentrum für Luft und Raumfahrt e.V.}, \orgaddress{\street{Lilienthalplatz 7}, \city{Braunschweig}, \postcode{38108}, \country{Germany}}}


\abstract{

The DESIRE project aims to test chameleon field theories as potential candidates for dark energy. The chameleon field is a light scalar field that is subject to screening mechanisms in dense environments making them hardly detectable. The project is designed to overcome this challenge. To this end, a specially designed source mass generates periodic gravitational and chameleon potentials. The design of the source mass allows for adjustment of the amplitude and periodicity of the gravitational potential while keeping the chameleon potential unchanged. The periodicity of the potentials makes them distinguishable from the environment and allows for resonant detection using multiloop atom interferometry under microgravity conditions.
}



\maketitle

\section{Introduction}\label{sec1}

In 1998, it was discovered that the expansion of the universe is accelerating \cite{riess_observational_1998,perlmutter_measurements_1999}. Since then, a plethora of theories have emerged to explain this phenomenon. 
In the standard model of cosmology, also referred to as $\Lambda$-CDM, this acceleration is taken into account by the cosmological constant which describes an energy density. However, there is no confirmed source for this energy density \cite{weinberg_cosmological_1989,martin_everything_2012}.
It has been proposed, in the frame of theories of modified gravity \cite{joyce_dark_2016}, to explain the cosmological constant by the existence of a scalar field that couples to all particles.
A screening mechanism is incorporated to ensure that the force associated with the scalar field complies with current solar system constraints and cosmological observations \cite{williams_progress_2004}.
The most popular of such scalar theories is the chameleon field theory \cite{khoury_chameleon_2004}, where the scalar field is modulated by its environment, meaning that the value of the scalar field depends on the local density of matter.
Such theories can be tested by investigating the attraction between objects.
Chameleon field theories can be tested in laboratory experiments, using classical sensors such as torsion balances \cite{adelberger_tests_2003}, or levitated force sensors \cite{yin_experiments_2022}.
In 2015, it was proposed to use atom interferometry to measure the gravitational interaction of atoms with a source mass and to test chameleon field theories, as atoms do not screen the chameleon field \cite{burrage_probing_2015}. Moreover, forces on atoms can be measured with high sensitivity by atom interferometry.
Such experiments with atoms have led to the exclusion of a broad range of parameters for chameleon field theory \cite{jaffe_testing_2017,sabulsky_experiment_2019,panda2024measuring}. A review of various experiments and the resulting constraints is provided in \cite{burrage_tests_2018}.
However, all these experiments were limited by the uncertainty on Newton’s constant $G=(6.67430 \pm 0.00015)\times 10^{-11}$~m$^3$kg$^{-1}$s$^{-2}$ \cite{NISTG}, limiting the accurate determination of the gravitational interaction between the source mass and the atoms. 

In this article, we present the DESIRE project which stands for Dark Energy Search by Interferometry in the Einstein-Elevator. This experiment is designed to overcome the limitations encountered in previous tests of chameleon field theory. 
A custom-designed source mass featuring a spatially periodic structure gives rise to the gravitational and hypothetical chameleon potential.
By modifying the mass distribution outside of the source mass, the periodicity and amplitude of the gravitational potential can be shifted with respect to the chameleon potential. 
The potential's periodicity enables resonant detection using multiloop atom interferometry \cite{chiow_multiloop_2018}. In order to reach a long interaction time and free inertial propagation to perform multiloop atom interferometry, the experiment will be performed in the Einstein-Elevator (EE) microgravity facility at Leibniz University Hannover \cite{lotz_einstein-elevator_2017}. This facility provides 4 s of microgravity up to 100 times per working day. For the atom interferometer setup, the MAIUS-1 payload, which demonstrated the first generation of a Bose-Einstein condensate (BEC) in space in 1.6 s \cite{becker_space-borne_2018}, is currently being modified and adapted to be mounted in the EE. 

The paper is organized as follows. We start in Section \ref{sec:atomInt_test} by introducing the chameleon field and deriving the associated acceleration felt by atoms. We also present the concept of the intended experimental test, including the design of the source mass and the resonant detection technique by multiloop atom interferometry. Section \ref{sec:Exclusion} discusses the expected parameter exclusion. In Section \ref{sec:ExpSetup} we detail the experimental setup. Finally, we conclude in Section \ref{sec:conclusion}.

\section{Atom-interferometric tests of the chameleon model}\label{sec:atomInt_test}

\subsection{The force associated with the chameleon field}

Atom interferometry is a great tool for testing the chameleon model, thanks to the diminishing thin-shell effect when single atoms are used as the force probe  \cite{burrage_probing_2015}. In this measurement scenario, the solution of the static equation of motion of the chameleon scalar field $\phi_{ch}$,
\begin{eqnarray}
    \nabla^2 \phi_{ch} &=& n_{ch} \Lambda^{4+n_{ch}}\phi_{ch}^{-n_{ch}-1}+\frac{\beta}{M_{Pl}} \rho,
\end{eqnarray}
depends on the density distribution $\rho$ of baryons, including the vacuum chamber, the supporting optomechanics, the atmosphere, etc, and on the model parameters $n_{ch},\Lambda,\beta=M_{Pl}/M$, where $M_{Pl}$ is the reduced Planck mass. $n_{ch}$ is an exponent that can either be a real number such that $n_{ch}>-1$ ($n_{ch} \neq 0$), or a negative even number $n_{ch}\leq -4$ \cite{burrage_tests_2018}. $\beta$ sets the coupling between the chameleon field and normal matter, and $\Lambda$ is the energy scale characterizing the chameleon field self-interaction.

The scalar field yields an acceleration $\vec{a}_{ch}$ on an atom:
\begin{eqnarray}\label{eqn:acc_chameleon}
    \vec{a}_{ch}&=&-\lambda_A \frac{\beta}{M_{Pl}} \vec{\nabla} \phi_{ch},
\end{eqnarray}
where $\lambda_A$ is the screening factor, $0\leq \lambda_A \leq 1$, which depends on $\beta$ and the local $\phi_{ch}$. $\lambda_A$ is usually extremely small but can be close to 1 for an atom \cite{burrage_probing_2015}. An effective potential can be defined such that 
\begin{eqnarray}
    \vec{a}_{ch}&=&-\vec{\nabla} V=-\vec{\nabla} \left(\lambda_A \frac{\beta}{M_{Pl}} \phi_{ch}\right),
\end{eqnarray}
for a constant or slowly varying $\lambda_A$, over the atom's trajectory.

\subsection{Multiloop interferometry, resonant detection}

\begin{figure}[H] 
    \centering
    \includegraphics[width=\textwidth]{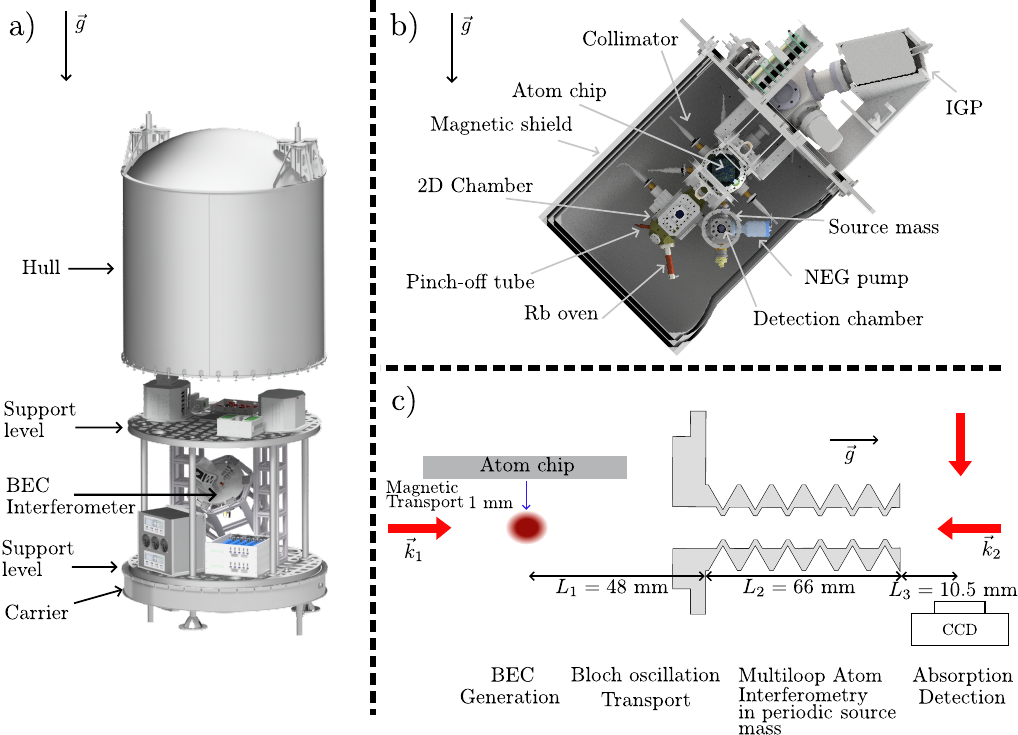}
    \caption{a) CAD drawing of the experimental setup placed in the inner capsule of the Einstein-Elevator formed by the hull and carrier. b) CAD drawing of the atom chip payload.${}^{87}$Rb atoms are loaded in a 3D MOT on the atom chip via the 2D MOT which transfer the atoms from the 2D chamber. The atoms are then transferred to a magnetic trap from which (RF) evaporative cooling leads to the formation of a BEC. c) Illustration of the experimental sequence in microgravity. The BEC is moved from the chip by nonlinear magnetic transport. After delta kick collimation (DKC), the BEC is transported to the source mass's entrance by Bloch oscillations. The atoms propagate through the source mass during the multiloop AI and are detected by absorption detection in the detection chamber.  }
    \label{fig:MicrogravitySetup}
\end{figure}

In contrast to other experiments based on accelerated or levitated atoms \cite{jaffe_testing_2017,sabulsky_experiment_2019,panda2024measuring}, the presented test will be based on a multiloop interferometer performed with a BEC while floating through a periodic source mass. During the experiment, the complete setup is in free fall for 4 s in an Einstein-Elevator as shown in Fig. \ref{fig:MicrogravitySetup}. 
The sequence starts with the production of a BEC on an atom chip. The BEC is then transported via Bloch oscillations to the entrance of the source mass through which multiloop atom interferometry will be performed. Finally, output port populations are detected by absorption detection. Details of the experiment during microgravity are given in the section \ref{sec:ExpSetup}. 
The multiloops are matched to the spatially varying force field.
After coherent splitting of the atomic wavepacket generated from a BEC, light pulses will toggle the velocity of the wavepackets in an alternating way. Consequently, in the space-time diagram, the wave packet trajectories form loops as depicted in Fig. \ref{fig:Principle_illustration}. The instances of the pulses, i.e., the velocity changes, are chosen such that they occur at the spatial period of the source mass, i.e., the potential of the hypothetical chameleon.
A multi-loop atom interferometer, as described in Ref.~\cite{chiow_multiloop_2018} and illustrated in Fig.~\ref{fig:Principle_illustration}, is constructed by a $\pi/2$-pulse to split the wave packet, multiple $\pi$-pulses to deflect the paths to form the loops, and a $\pi/2$-pulse to recombine and interfere the paths at the end. In the following, we show that the differential phase $\phi = \varphi_I - \varphi_{II}$ of a pair of $N$-loop atom interferometers, where the loops are synchronized with the evenly spaced crests and troughs of a weak periodic potential $V_p$, is given by $\phi = 2 N T \delta V_p m / \hbar$. Here, $\delta V_p$ is the crest-to-trough difference of the potential, $T$ is the pulse separation time between $\pi/2-$ and $\pi-$ pulses, $m$ is the mass of the atom, and $\hbar$ is the reduced Planck constant.

\subsection{Sensitivity estimate of multiloop interferometers}

\begin{figure}[htbp]
\centering
\includegraphics[]{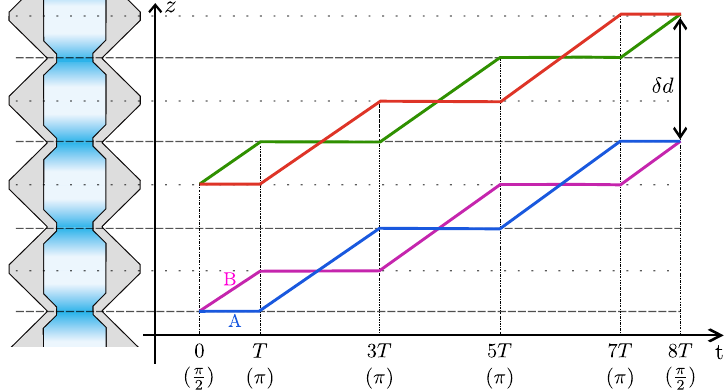}
\caption{Space-time diagram of dual 4-loop AIs. In the lower AI (blue and magenta), the atom starts at rest at a potential crest. The wave packet is split into two after the first beam splitter pulse leading to the two wave packets A and B. The blue gradient illustrates the chameleon potential in the source mass, it is maximum in the regions of minimum inner radius. We see on this illustration that the wave packet A is always at rest at potential crest while the wave packet B is at rest at potential through.}
\label{fig:Principle_illustration}
\end{figure}

We adapt the path integral method \cite{storey1994feynman} to calculate the phase $\varphi$ of one multiloop atom interferometer, which can be divided into three parts: the propagation phase $\varphi_\textrm {prop}$, the laser phase $\varphi_\textrm {laser}$, and the separation phase $\varphi_\textrm {sep}$.
The propagation phase is the difference of the action integral:
\begin{eqnarray}
\varphi_\textrm{prop}&=&\frac{1}{\hbar}\int_B \frac{m}{2} v^2 -V(x)\ dt\nonumber\\
&&-\frac{1}{\hbar}\int_A \frac{m}{2} v^2 -V(x)\ dt,
\end{eqnarray}
where $v$ is the velocity, and $V=V_p+V_b$ is the total potential energy that the atom experiences. $ V_b$ is a background potential.
The integrals are taken along the classical trajectories $A$ and $B$ of the two paths.
The duration of atom-photon interactions during the laser pulses is assumed to be infinitesimal compared to the pulse separation time $T$ and considered instantaneous.

Assuming that, in the absence of the weak periodic potential $V_p$, the background potential $V_b$ is flat and that the atom starts from rest, each path spends an equal amount of time at $v=0$ and at $v=2nv_r$. Here, $v_r$ is the recoil velocity and $2n$ is the number of photon momentum interchanged during an atom-light interaction.
Therefore, both the kinetic and the potential energy integrals are identical and do not contribute to $\varphi_\textrm{prop}$.
Now, treating $V_p$ as a perturbation so that the trajectories are not affected to the leading order, $\varphi_\textrm{prop}$ can be rewritten as
\begin{eqnarray}
\varphi_\textrm{prop}&=&-\frac{1}{\hbar}\left(\int_B  V_p(x)\ dt-\int_A  V_p(x)\ dt\right),
\end{eqnarray}
where the integrals are taken along the unperturbed trajectories $A$ and $B$.

As illustrated in Fig. \ref{fig:Principle_illustration} for the lower (blue and magenta) AI, if the atom starts at rest on a potential crest, half of the wave packet ($B$) goes toward the next trough at $v=2nv_r$ while the other half ($A$) stays at the crest after the $\pi/2$-pulse.
At time $T$, when the first half ($B$) reaches the trough, a $\pi$-pulse is applied so that the first half is stationary while $A$ starts to move at $v=2nv_r$.
After time $2T$, where the moving wave packet $A$ has passed the trough and reached another crest, a second $\pi$-pulse is applied.
$A$ then becomes stationary at the crest while $B$ starts to move away from the trough and towards the crest where $A$ is.
After another $2T$, where $B$ has crossed $A$ and the crest and reached another trough, the next $\pi$-pulse is applied.
$A$ then starts to move towards $B$ and the next crest while $B$ stays stationary at the trough.
The process continues until after an even number of loops $N$ are formed, a $\pi/2$-pulse is applied at a crest where $A$ and $B$ overlap.
$A$ spends half of the time at the crest and the other half moving up and down $V_p$, while $B$ spends half of the time at the trough and the other half moving down and up $V_p$.
Due to symmetry, $A$ and $B$ spend an equal amount of time going up and going down $V_p$, thus the potential energy integrals during transients are identical and cancel.
Hence, $\varphi_\textrm{prop}$ is equivalent to $A$ spending half of the time at the crest and $B$ spending half of the time at the trough, resulting in
\begin{eqnarray}
\varphi_\textrm{prop}&=&-\frac{1}{\hbar}\left(\int_B  V_p(x)\ dt-\int_A  V_p(x)\ dt\right)\nonumber\\
&=&\frac{1}{\hbar}\left(\int_{A \textrm{ at crest}}  V_p(x)\ dt+\int_{A \textrm{ transient}}  V_p(x)\ dt\right.\nonumber\\
&&\left.-\int_{B \textrm{ at trough}}  V_p(x)\ dt-\int_{B \textrm{ transient}}  V_p(x)\ dt\right)\nonumber\\
 &=&\frac{m}{\hbar} NT \delta V_p,\label{eq:prop}
\end{eqnarray}
where the total time for the $N$-loop interferometer is $2NT$, and $m\ \delta V_p$ is the crest-to-trough energy difference of $V_p$.
Note that the result is independent of the exact shape of the potential, which relaxes the knowledge of $V_p$ and eases the optimization of the source mass.

The separation phase $\varphi_\textrm{sep}$ arises when the two classical trajectories do not coincide at the last $\pi/2$-pulse, which is caused by position-dependent forces such as gravity gradients.
Under the assumption of a flat background potential, the trajectories coincide after the pulse sequence, and thus $\varphi_\textrm{sep}=0$.

$\varphi_\textrm{laser}$ is the phase acquired during atom-photon interactions.
The wave packet acquires (deducts) the effective phase of the beamsplitter when gaining (losing) the photon momenta, and no phase change otherwise.
Since both paths experience the same laser pulses, it is straightforward to conclude that $\varphi_\textrm{laser}=\varphi_{\pi/2_1}-2\varphi_{\pi_1}+2\varphi_{\pi_2}-\cdots+2\varphi_{\pi_N}-\varphi_{\pi/2_2}$, where $\varphi_{\pi/2_{1,2}}$ are the laser phases of the beamsplitters, and $\varphi_{\pi_{i}}$ is the laser phase of the $i$-th $\pi$-pulse.
Thus, the total phase of the $N$-loop interferometer is $\varphi=NT\delta V_p m/\hbar+\varphi_{\textrm{laser}}$.

Several factors contribute to $\varphi_\textrm{laser}$, such as the phase set at the source of light, differential optical phase delay between frequency components, and vibrational disturbances to optical paths.
In sensitive interferometers, vibrational noise is the dominating factor that smears the fringe via $\varphi_{\textrm{laser}}$.
Instead of implementing vibration isolation mechanisms or compensation schemes, differential atom interferometers are commonly used in certain applications, where two displaced atom interferometers are simultaneously driven by the same set of laser pulses.
It is demonstrated that the suppression of common mode noise in $\varphi_\textrm{laser}$ between interferometers exceeds 100 dB \cite{mcguirk2002sensitive}.
In the proposed scheme, two $N$-loop atom interferometers are addressed simultaneously by the same laser pulses.
One interferometer starts at rest from a potential crest with an interferometer phase $\varphi_{I}=NT\delta V_p m/\hbar+\varphi_\textrm{laser}$, as described earlier.
The other interferometer also starts at rest but from a potential trough, whose phase is thus $\varphi_{II}=-NT\delta V_p m/\hbar+\varphi_\textrm{laser}$.
The resulting differential phase of the interferometer pair is
\begin{equation}\label{eqn:diff_phase}
 \phi=\varphi_{I}-\varphi_{II}=2NT \delta V_p m/\hbar,   
\end{equation}
which is insensitive to vibrational noise and other laser noises.

Note that $m/\hbar$ is intrinsic to the atomic species used in the interferometer, and $2NT$ is the total interferometer time.
Equation (\ref{eqn:diff_phase}) suggests that if the interferometer time is limited by the apparatus, there is a tradeoff for sensitivity between the number of loops $N$ and the loop time $T$. On the other hand, large-momentum-transfer beamsplitters (LMTs), which affect $n$, do not improve the sensitivity directly, although LMTs enable shorter $T$ for the same periodicity.
The sensitivity increases linearly with the atomic mass, the total interferometer time, and the periodic potential variation.
To put numbers in perspective, $2NT=1.9$~s as available on the Einstein-Elevator and $m=87$~amu ($^{87}$Rb), then $\phi=2.6\times 10^{9} \delta V_p$.
For a pair of shot-noise-limited atom interferometer, the differential phase noise is $\delta \phi=2/ (C\sqrt{N_{at}})$, where $C$ is the contrast and $N_{at}$ is the atom number in each interferometer. For $N_{at}=5\times 10^4$, $C=0.5$, the uncertainty in $\delta V_p$ is $\delta \phi/(2.6\times 10^9)=6.87\times 10^{-12}$ m$^2$/s$^2$ per experiment run.

\subsection{The source mass} 
To take full advantage of the MAIUS-1 payload and the EE facility, the source mass geometry is optimized for the available free-fall time, spatial constraints of the MAIUS-1 package, and fabrication technology.
Additive manufacturing (AM) of metal is chosen as the preferred manufacturing approach for the source mass, instead of conventional machining or welding of sub-assemblies. Thermal deformation due to welding, material gain and loss during welding, ultra-high vacuum (UHV) compatibility, and assembly geometry tolerance are risks for the project. 

AM has been demonstrated to form UHV enclosures using aluminum or titanium alloys \cite{vovrosh_additive_2018}. To obtain the best cylindrical symmetry, the source mass is to be manufactured along the symmetry axis. This constrains the recessed surface's overhang angle of less than 45$^{\circ}$. Also, features are limited to no smaller than 0.1 mm, in addition to the limited choice of materials. 
The source mass is made of titanium alloy Ti-6Al-4V, for its well-understood AM properties, and the thermal compatibility to the MAIUS-1 titanium vacuum enclosure. 

The outer radius of the source mass $r_2=11$~mm is limited by the clearance between magnetic coils in the MAIUS apparatus. 
The other dimensions of the source mass are determined as follows. For a $2n\hbar k$, $N$-loop interferometer starting from rest, to be detailed in the next section, the total spatial extent is $L_{total}=2 n N v_r T$, and the total duration is $T_{total}=2NT$, with $v_r\simeq 5.88$~mm/s and $k$ is the interferometry laser wavenumber. The spatial periodicity $L$ that this interferometer is sensitive to is the length of two loops, $L=4n v_r T$, therefore, $N$ has to be an even number. 
The length of the source mass should be short enough that atom interferometers using no greater than $4\hbar k$ beamsplitters can traverse in 1.9~s (available time for the multiloop atom interferometry in microgravity). $4\hbar k$ momentum transfer is chosen to ensure good diffraction efficiency and manageable optical power requirement. This leads to the constraints $T_{total}\le1.9$~s, $n\le2$, and $N\ge4$. This results in the length of each segment $L=2L_{total}/N=2T_{total}n v_r/N \le 11.2$~mm. We choose $L=11$~mm to have a slight margin.

\begin{figure}[htbp]
\centering
\includegraphics[width=0.8\textwidth]{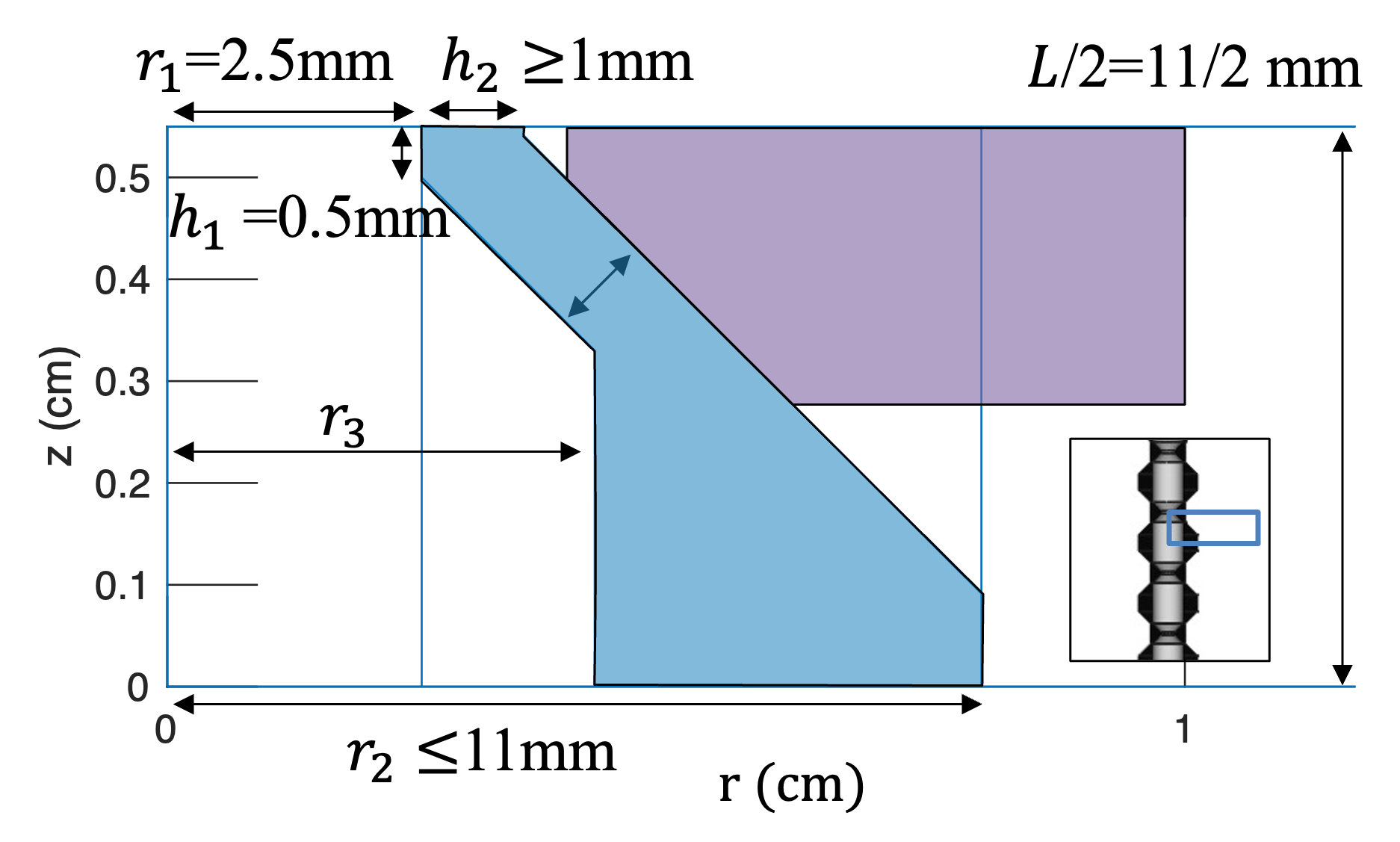}
\caption{Simulation configuration. The simulation is set up in the cylindrical coordinates $(r,z)$. The colored region demotes the body of the structure made of titanium. To the left of the body is the ultra-high vacuum, while to the right is the nominal atmosphere. The range of $z$ is half of the periodicity $L/2$, while the range of $r$ is chosen sufficiently large that the field variation along axis $(r=0)$ inside UHV is not affected by the larger range. The purple polygon represents the lower half of an additional tantalum ring for generating periodic gravity potential. The inset shows an infinite periodic structure and the domain of simulation.}
\label{fig:simulation_setup}
\end{figure}

The source mass is first made via the AM process, followed by chemical etching, heat treatment, and hot isostatic pressing. Finally, conventional machining is applied to the adapter plate surfaces and the lip structure, to ensure sufficient surface quality for vacuum sealing and welding, respectively.

Gravitational and chameleon fields are determined numerically using MATLAB finite element method~\cite{chiow_multiloop_2018}. The source mass is parameterized as depicted in Fig.~{\ref{fig:simulation_setup}}. Due to the cylindrical symmetry, the periodicity, and the mirror symmetry within each section, the domain of simulation is half of the periodicity with the Neumann boundary condition along the axial dimension $z$ and at $r=0$. The boundary condition at the maximum $r$ is either set to 0 for gravity or to the vacuum expectation value (i.e., $\phi_{ch}=(\frac{n_{ch} M_{Pl}\Lambda^{n_{ch}+4}}{\beta\rho })^{\frac{1}{n_{ch}+1}}$ for a local baryon density $\rho$) of the dark energy field. This boundary condition does not affect the resulting field variation at $r=0$, due to the dark energy models' screening mechanism and the gravity field's linearity.

\begin{figure}[htbp]
\centering
\includegraphics[width=0.9\textwidth]{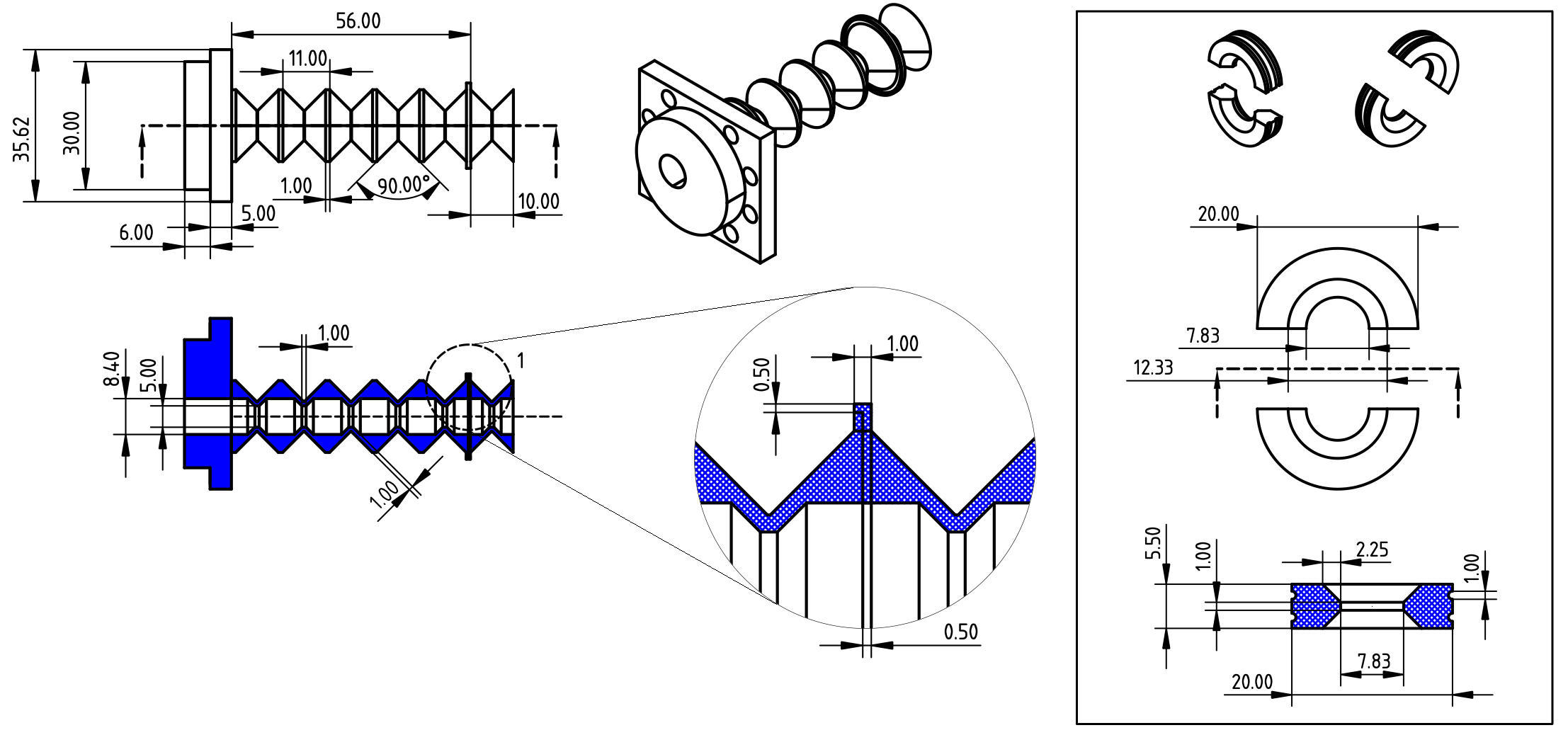}
\caption{Sketch of the source mass. The through-hole is not centered on the adapter plate to accommodate the anticipated cloud position from the atom chip. The inset depicts the sketches for the tantalum half-ring that can be attached to the source mass.}
\label{fig:sourcemass}
\end{figure}

The simulation results of the chameleon model for $n_{ch}=1$, see~\cite{chiow_multiloop_2018}, show that the chameleon field variation is insensitive to $r_2$, as expected, but increases as $r_3$ is increased. Thus, the length of the narrow throughhole, $2h_1$, is set to the smallest allowable feature size of 1~mm. The values of $h_2=1$~mm and $r_3=4.2$~mm are determined by minimizing both the gravity potential difference between $z=0$ and $z=L/2$ and the peak-to-peak variation at $r=0$. Figure~\ref{fig:sourcemass} shows the source mass design with an adapter plate to the MAIUS-A vacuum chamber and a lip for welding to a detection chamber made of titanium.

To validate the multi-loop atom interferometers, the periodicity and amplitude of the gravity signal are modulated by additional tantalum rings that can be attached to the source mass structure externally. 

\begin{figure}[htbp]
\centering
\includegraphics[width=\textwidth]{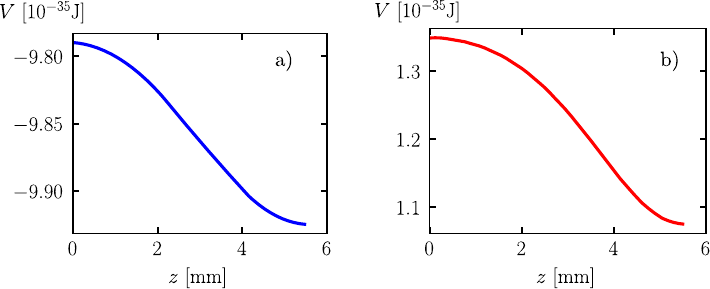}
\caption{a) Gravity and b) chameleon potentials with the tantalum rings attached to the source mass over half the periodicity of the source masse. The gravity and chameleon potentials have the same spatial periodicity. Also, the amplitude variations of the gravitational potential are similar to the chameleon one.}
\label{fig:withTrings}
\end{figure}

When the tantalum rings are attached to the source mass, the gravity signal has similar periodicity and amplitude variations as the chameleon one, see Fig.~\ref{fig:withTrings}.
The acceleration due to the gravity potential when the tantalum rings are attached is expected to be one order of magnitude smaller than the acceleration due to the chameleon potential. The detection of the gravity potential when the tantalum rings are attached enables the demonstration of the measurement scheme.

\begin{figure}[H]
\centering
\includegraphics[width=\textwidth]{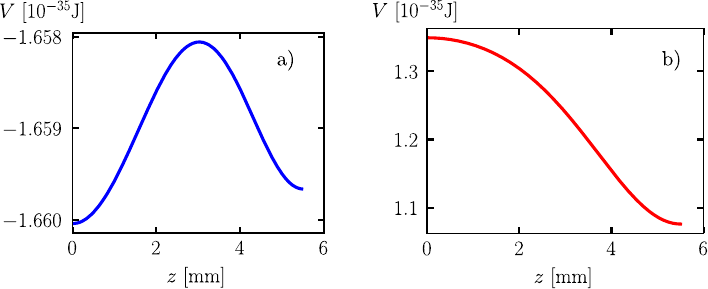}
\caption{a) Gravity and b) chameleon potentials without the tantalum rings attached to the source mass over half the periodicity of the source masse. The gravity potential has a spatial periodicity two times smaller than the chameleon potential. Also, the amplitude variations of the gravitational potential are smaller than the chameleon one.}
\label{fig:withoutTrings}
\end{figure}

Without the tantalum rings, the gravity signal has a periodicity two times larger than the chameleon and much smaller gradients, as shown in Fig.~\ref{fig:withoutTrings}.

Tantalum is chosen for its AM compatibility and higher density of 16.7~g/cm$^3$ than titanium 4.51~g/cm$^3$. The thickness of the ring is $2.75$~mm, half of the periodicity, and the outer radius is set to $10$~mm. Each ring is composed of two half rings with two grooves on the outer surface, to allow attachment to the source mass using wire fastening, as depicted in the inset of Fig.~\ref{fig:sourcemass}.

\subsection{Background potential suppression in multiloop interferometers}\label{sec:suppression}
For even $N$, an $N$-loop atom interferometer can be considered as a series of two-loop interferometers, each of which is a gradiometer.
The phase of the interferometer is $N/2$ times the gradient-induced phase of each gradiometer.
Let the background potential be expressed as:
\begin{eqnarray}
V_b&=&V_{b0}+V_{b1} z+\frac{1}{2}V_{b2} z^2+\frac{1}{6}V_{b3} z^3+\cdots,
\end{eqnarray} where $V_{b1}$ corresponds to linear acceleration, $V_{b2}$ to gradient, and so on.
$V_b$ can include effects due to rotation, Zeeman shifts, and pseudo forces.
To the leading orders, the phase of a $T-2T-T$ two-loop gradiometer starting at position $z_0$ with an initial velocity $v_{0}$ is:
\begin{eqnarray}
\varphi^{(2)}(z_0)&=&-\bar{v}k_\textrm{e} T^3\left[V_{b2}+ (z_0+\bar{v} T)V_{b3} +\cdots\right],
\end{eqnarray}where $\bar{v}=2nv_r+2v_0\equiv v_{nr}+2v_0$ and $k_\textrm{e}=2nk$.
Note that $\bar{v}T$ is the mean distance that the atom traverses after one loop.
For two $N$-loop interferometers that respectively start at $z_0$ and $z_0+\delta d$, the differential phase can be readily obtained:
\begin{eqnarray}
\phi(z_0,\delta d)&=&\sum_{l=0}^{N/2-1}\varphi^{(2)}(z_0+2\bar{v} T l+\delta d)-\varphi^{(2)}(z_0+2\bar{v} T l)\nonumber\\
&=&-\bar{v}k_\textrm{e} T^3\left[ \frac{1}{2}\delta d \ N V_{b3} +\cdots\right].\label{eq:eq1}
\end{eqnarray}
Evidently, $\phi(z_0,\delta d)$ is immune to quadratic potential $V_{b2}$, significantly eliminating the influence of the spatial variation of environmental mass distribution, such as the apparatus, the platform, and the celestial bodies. 
Intuitively, dual $N$-loop interferometers are $N/2$ repetitions of doubly differential Mach-Zehnder interferometers.
Thus, systematic effects in a Mach-Zehnder interferometer would appear in $\phi(z_0,\delta d)$ only if they are quadratic to displacement, which would occur at high-order iterative expansion and are negligibly small.

To illustrate the effect of spatial synchronization between the interferometers and a periodic potential, the phase due to the perturbing potential is calculated along classical trajectories in a flat background potential.
Let us consider a sinusoidal potential $V(z)=V_K \cos (Kz)$.
The phase of a single loop is
\begin{eqnarray}
\phi^{(1)}(z_0)&=&-\frac{m}{\hbar}\frac{v_{nr}V_K}{K v_0  (v_0+v_{nr})}\left[\sin(K z_0)-\sin(K (z_0+v_0 T))\right.\nonumber\\
&&-\sin(K (z_0+(v_0+v_{nr}) T))\nonumber\\
&&\left.+\sin(K (z_0+(2v_0+v_{nr}) T)) \right].
\end{eqnarray}
The phase of $N$ loops can be expressed as
\begin{eqnarray}
\phi(z_0)&=&\sum_{l=0}^{N-1}(-1)^l\phi^{(1)}(z_0+l \bar{v}T)\nonumber\\
&=&-\frac{m}{\hbar}\frac{v_{nr}V_K T}{ v_0+v_{nr}}\left[\frac{\sin(K z_0)-\sin(K (z_0+v_0 T))}{K v_0 T}\right.\nonumber\\
&&\left.-\sum_{l=0}^{N-2}(-1)^l\frac{\sin (K(z_0+(l+1) v_{nr}T+(2l+1)v_0 T))-\sin (K(z_0+(l+1) v_{nr}T+(2l+3)v_0T))}{K v_0 T}\right.\nonumber\\
&&\left.+\frac{\sin(K (z_0+N v_{nr}T+(2N-1)v_0T))-\sin(K (z_0+N v_{nr}T+2Nv_0 T))}{K v_0 T}\right]\nonumber\\
&=&\frac{m}{\hbar}\frac{v_{nr}V_K T}{ v_0+v_{nr}}\textrm{sinc}\!\left(\frac{K v_0T}{2}\right)\left[\cos(K (z_0+\frac{v_0T}{2}))\right.\nonumber\\
&&\left.-\sum_{l=0}^{N-2}(-1)^l \left(\cos(K (z_0+((l+1) \bar{v}+\frac{1}{2}v_0)T))+\cos(K (z_0+((l+1) \bar{v}-\frac{1}{2}v_0)T))\right)\right.\nonumber\\
&&\left.+\cos(K (z_0+(N \bar{v}-\frac{1}{2}v_0)T))\right].\label{eq:phi}
\end{eqnarray}
Since $\phi(z_0)$ is proportional to the sinc function of $v_0$, nonzero initial velocity or velocity spread of the ensemble decreases the signal size.
In the limit of small initial velocity $K v_0 T/2\equiv a\ll1$, the above expression reduces to
\begin{eqnarray}
\phi(z_0)&=&\frac{m}{\hbar} T\left(1-\frac{a^2}{6}\right)\left(1-\frac{a^2}{2}\right)\left[V( z_0)+2\left(\sum_{l=1}^{N-1}(-1)^l V(z_0+l v_{nr}T)\right)+V(z_0+N  v_{nr}T)\right]\nonumber\\
&&+\frac{m}{\hbar} T \left(1-\frac{a^2}{6}\right)a \left[-V\left(z_0-\frac{\pi}{2K}\right)+V\left(z_0+N  v_{nr}T-\frac{\pi}{2K}\right)\right].\label{eq:phi2}
\end{eqnarray}
If $V(z)=V(z+2  v_{nr}T)$, i.e., $K 2 v_{nr}T$ is an integer-multiple of $2\pi$, then 
\begin{equation}\phi(z_0)= \frac{m}{\hbar}NT\left(V(z_0)-V(z_0+\bar{v}T)\right)+\mathcal{O}(a^2),\label{eq:phi3}
\end{equation} which recovers Eq.~(\ref{eq:prop}), which represents a specific alignment of two $N$-loop interferometers to $V(z)$.
However, if $V(z)\ne V(z+2  v_{nr}T)$, the sum of all loop contributions is reduced. 
For an infinite number of loops, the sum is zero; for finite $N$, the suppression of the maximal sum due to a period mismatch $\delta p= 2\pi-K 2 v_{nr}T$ can be estimated as 
\begin{eqnarray}
\frac{\phi(z_0,\delta d;\delta p)}{ \phi(z_0,\delta d)}&=&\frac{1}{N}\sum_{l=0}^{N-1}\cos l \delta p\nonumber\\
&=&\frac{1}{N}\frac{\sin\frac{N \delta p}{2}}{\sin\frac{ \delta p}{2}}\cos\frac{(N-1)\delta p}{2}\nonumber\\
&=&\frac{\textrm{sinc}\!\left(\frac{N \delta p}{2}\right)}{\textrm{sinc}\!\left(\frac{ \delta p}{2}\right)}\cos\frac{(N-1)\delta p}{2}\nonumber\\
&\simeq&1-\frac{N^2}{6}\delta p^2,\textrm{     for small $\delta p$}.\label{eq:dp}
\end{eqnarray}
Thus, the spatial frequency selectivity is proportional to $N^2$.

More generally, for a periodic but non-sinusoidal potential, contributions from all spectral components add independently to $\phi$, due to the perturbative nature of the treatment.
Specifically, $\phi$ is the sum of potential components at the primary frequency of the field mass and its harmonics.
Other features in the potential, whether periodic or not, are suppressed according to Eq.~(\ref{eq:dp}).

Experimentally, $\delta p$ is to be varied by changing $T$, and the resulting phase profile is to be fit to Eq.~(\ref{eq:dp}).
This $\delta p$ survey not only ensures proper alignment of the periodicity and thus accurate gravity measurements but also serves to suppress systematic errors, since $\phi(z_0,\delta d)$ may contain terms that are constant or slowly varying in $z_0$ (e.g., due to the cubic potential $V_{b3}$ in Eq.~(\ref{eq:eq1})).
Moreover, the starting point $z_0$ can also be explored experimentally. In this scenario, the measured $\phi(z_0)$ will be projected to the expected signature of $V(z_0)-V(z_0+\bar{v}T)$, further suppressing systematic errors and providing better accuracy for non-zero measurements.

\subsection{Estimation of the different surface-related potential}
Inside the source mass, the Casimir-Polder potential has the same periodicity as the chameleon potential but is expected to be negligible. To give orders of magnitude, we consider a ${}^{87}$Rb atom in its ground state in front of an infinitely perfect conductor surface at temperature $T=300~\text{K}$. For atom-surface distance larger than the thermal photon wavelength (justified by the source mass dimensions), the Casimir-Polder potential in thermal equilibrium writes \cite{buhmann_dispersion_2013,buhmann_dispersion_2013-1}
\begin{equation}\label{atom-surface}
    V_{CP}(l)= -\frac{k_B T \alpha(0)}{16\pi \epsilon_0 l^3}.
\end{equation}
Here, $\alpha(0) \approx 5.25\times 10^{-39}\ \text{J(m/V)$^2$}$ is the static polarizability of ${}^{87}$Rb ground state \cite{holmgren_absolute_2010}. $l$ is the atom-surface distance, $k_B$ the Boltzmann's constant and $\epsilon_0$ the vacuum permittivity. Thus, in our simple model, the Casimir-Polder potential for an atom located 2.5~mm away from the wall is of the order of $\left|V_{CP}(l=2.5\ \text{mm})\right|\approx 3.1 \times 10^{-42}\ \text{J}$. The acceleration due to the potential (\ref{atom-surface}) for a ${}^{87}$Rb  atom of mass $m$ is given by
\begin{equation}
    \left|a_{CP}(l)\right| = \frac{3 k_B T \alpha(0)}{16\pi \epsilon_0 m l^4},
\end{equation}
which in our case is of the order of $a_{CP}(l=2.5~\text{mm})\approx 2.6\times 10^{-14}~\text{m/s$^2$}$. Therefore, the acceleration due to the Casimir-Polder potential is expected to be negligible compared to the acceleration due to the hypothetical chameleon potential. 

For non-thermal equilibrium, it has been shown in previous precision experiments that black-body radiation can have a measurable effect on atom interferometers close to surfaces \cite{haslinger_attractive_2018}. However, it is unlikely that the temperature distribution of the source mass exhibits a spatial periodicity matching the source mass periodicity. Hence, such black-body radiation potential would not be picked up by the multiloop atom interferometer.

\section{Expected exclusion}\label{sec:Exclusion}

As described in Section~\ref{sec:atomInt_test}, a null result of extra force constrains the allowable parameter space of the model under evaluation.
With the expected per-run $\delta V_p$ uncertainty of $6.87\times10^{-12}$~m$^2$/s$^2$, the constraint is equivalent to a contour line representing parameter combinations that would produce a  periodic potential variation of $\delta V_p$. 
While in the literature the contour line is usually established by numerical simulation on a single data point followed by arguments of force scaling dependence on parameters, we note that the spatial extent of atom interferometers is not explicitly discussed while the length scale of the chameleon force changes dramatically over large parameter ranges.
In the DESIRE experiment, it is established in Section~\ref{sec:atomInt_test} that $\delta V_p$ is indeed what will be measured for all length scales of forces of interest. 
Furthermore, the contour line is assembled by parameter pairs, each of which is numerically verified to generate $\delta V_p$ equivalent to the assumed experimental uncertainty. 
The inclusion of the scaling factor $\lambda_A$ in $V$ also makes our approach robust in situations where $\lambda_A$ varies significantly in the region of interest.

\begin{figure}[htbp]
\centering
\includegraphics[width=0.65\textwidth]{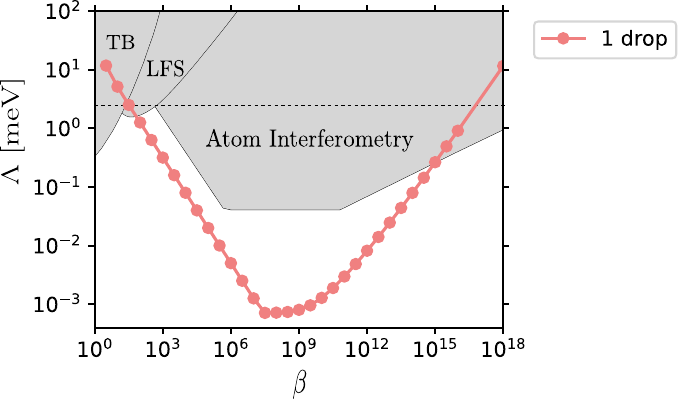}
\caption{Chameleon exclusion plot for $n_{ch}=1$. Points indicate numerically validated parameter pairs, while the line is to guide the eye. The region above the curve is excluded from the expected measurement sensitivity. TB illustrates the parameter excluded by torsion balance experiments \cite{burrage_tests_2018} and LFS by levitated force sensor \cite{yin_experiments_2022}. Atom interferometry refers to \cite{jaffe_testing_2017}.}
\label{fig:cha1}
\end{figure}

Figure~\ref{fig:cha1} shows the expected exclusion region for chameleon with $n_{ch}=1$ based on the expected per-run sensitivity. It is clear that the shape of the contour is different from other atom interferometer experiments. There is a clear advantage by up to 20~dB near $\beta=10^8$ of using structured source mass compared to a conventional spherical vacuum chamber. However, the scaling seems to be unfavorable in our approach for $\beta>10^{10}$, where the chameleon force is extremely short-ranged. It is unclear if analyses from the literature consider that the interferometer might be orders of magnitude larger than the characteristic length scale of the field.

\begin{figure}[H]
\centering
\includegraphics[]{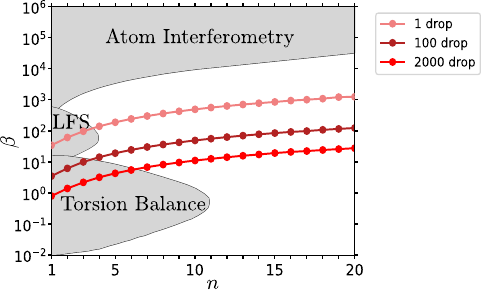}
\caption{Chameleon exclusion plot for $\Lambda=2.4$~meV. Points indicate numerically validated parameter pairs, while the line is to guide the eye. The region above each curve is excluded from the expected measurement sensitivity. Gray regions represent already excluded parameters.}
\label{fig:cha2}
\end{figure}

\begin{figure}[H]
\centering
\includegraphics[]{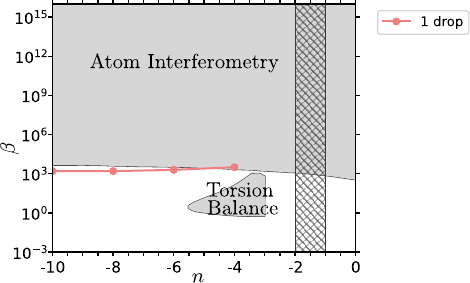}
\caption{Chameleon exclusion plot for $\Lambda=2.4$~meV for negative $n$ values adapted from \cite{burrage_tests_2018}. Points indicate numerically validated parameter pairs, while the line is to guide the eye (only negative even integers are compatible with the chameleon mechanism). The region above the curve is excluded from the expected measurement sensitivity. Gray regions represent already excluded parameters, in the hashed area the model is not chameleon.}
\label{fig:cha3}
\end{figure}

Exclusion at $\Lambda=2.4$~meV for different $n_{ch}>0$ is plotted in Fig.~\ref{fig:cha2}, where constraints from different numbers of experiment runs are included to show how the result will improve with extended data collection. There is also interest in negative $n_{ch}$, as plotted in Fig.~\ref{fig:cha3}. The numerical convergence however is poor for $n_{ch}<0$.

\section{Experimental setup to test chameleon field theory}\label{sec:ExpSetup}
For the DESIRE experimental project, the specially designed source mass generating both the gravitational potential and the expected chameleon potential presented in section \ref{sec:atomInt_test} has been added to the MAIUS-A payload \cite{grosse_design_2016}. The payload is adapted to be mounted on support levels to be placed in the EE facility as illustrated in Fig.~\ref{fig:MicrogravitySetup} a).

The Einstein-Elevator facility is located at the HiTec Institute in Hannover. The 27$\,$m high tower offers up to 4.0$\,$s of microgravity \cite{lotz_einstein-elevator_2017}. Between flights, a 4-minute break is necessary to cool down the brakes and to ensure the reliable operation of the Einstein-Elevator. 
Thus, a total of 100 flights per working day is possible.
To create an environment at atmospheric pressure for all subsystems, an additional inner capsule (pressure hull and carrier) is provided so that only the area between the gondola and the inner capsule is under vacuum (1$\,$Pa). This also ensures acoustic decoupling to further reduce vibrations.
The hull weighs 120$\,$kg and offers an interior space of 167$\,$cm (diameter) $\times$ 200$\,$cm (height). It is mounted on a base containing electronics, sensors, control systems, and a water pump for water cooling. The weight of the base part is 180$\,$kg (without components).
Parameters such as acceleration, microgravity quality, rotation, and magnetic field changes can be monitored during flights. For the magnetic field, a maximum fluctuation of 10$\,\mu$T has been measured along the entire trajectory, with rotations (order of magnitude in mrad per s) compensated to a few $\mu$rad per second by reaction wheels within the system.
During the ground time, a total of 1$\,$kW cooling power is supplied to the experiment via water cooling. In addition, the idle time between runs is used to recharge the supercapacitors and to carry out calibration measurements.
Two support levels will be used, each weighting $50~\text{kg}$, leaving 600$\,$kg for the experimental setup including all support equipment as the total permissible weight in the nacelle is 1000$\,$kg.

To mitigate rotation effects, the different modules will be placed such that the center of mass of the overall setup matches the midpoint of the straight line  between the center of the atom chip and the center of the detection chamber. The center of mass can be set with $\mu$m accuracy using additional mass elements. 
Furthermore, as illustrated in Fig.~\ref{fig:MicrogravitySetup}, the vacuum chamber will be held such that the symmetry axis of the source mass is aligned with Earth's gravitational axis $\vec{g}$ to perform experiments on ground.

\begin{figure}[htbp]
\includegraphics[]{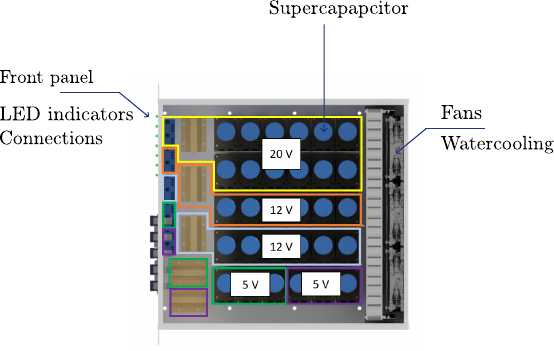}
\caption{Illustration of a supercapacitor power supply rack from above.}\label{fig:supercap_rack}
\end{figure}

During flights, the main control electronics, used during the MAIUS-1 mission, is powered only by super-capacitors (HY-CAP 500F 3V) connected in series for each required voltages ($20\,$V, $13\,$V, $10\,$V, $7\,$V, $5\,$V, $3.3\,$V, $-7\,$V, $-13\,$V and $-20\,$V) and build in three $19\,$" racks, see Fig. \ref{fig:supercap_rack}. Each super-capacitor is mounted on a safety board such that they are limited to $2.5\,$V and therefore have a longer lifespan.
Between flights, the super-capacitors are charged via R\&S HMP4040 power supply units. After each $4\,$s flight, the super-capacitors will be discharged at maximum by $5\,$\%, to maintain minimum voltage to keep the experiment running. All super-capacitors are fully charged during the 4 minutes between flights to maintain the same conditions for each flight.

\subsection{Details of experimental Sequence during microgravity}
To save microgravity time, the 3D MOT will be loaded before the flight. The maximum force in the 3D MOT is given by the scattering force at resonance
\begin{equation}
    \vec{F}_{max}=\hbar \vec{k} \frac{\Gamma}{2} \frac{I/I_{sat}}{1+I/I_{sat}},
\end{equation}
where $I$ is the laser beam intensity, $I_{sat}\approx3.576\ \text{mW/cm²}$ is the saturation intensity \cite{SteckRb}, $\vec{k}$ is the laser wave vector, and $\Gamma=38.11\ $~MHz is the natural linewidth of ${}^{87}$Rb D$_{2}$ line. The 3D MOT laser beams have Gaussian intensity profile and a waist radius $\approx 5.5~\text{mm}$, and the total power per beam is 20.0 mW leading to a peak intensity $I_0\approx42\ \text{mW/cm²}$.
Taking into account that the laser beams of the 3D MOT have a 45° angle with respect to the Earth's gravitational axis, and since the atoms absorb light from one of the laser beams at a time, the maximum force along the acceleration axis is given by
\begin{equation}\label{eqn:f_max}
   \vec{F}_{max}\cdot\frac{\vec{g}}{\vec{||g||}}=\hbar k \frac{\Gamma}{2} \frac{I/I_{sat}}{1+I/I_{sat}}\cos(\pi/4)\approx 1.04\times10^{-20} \ \text{N}, 
\end{equation}
where $\vec{g}$ is earth gravitational acceleration vector. We can now compare (\ref{eqn:f_max}) to the inertial force due to the acceleration of the EE carrier which can reach up to 5g
\begin{equation}
    5gm/(\vec{F_{max}}\cdot\frac{\vec{g}}{\vec{||g||}})\approx 7\times10^{-4}.
\end{equation}
\noindent This indicates that the maximum trapping force in the 3D MOT is much larger than the inertial forces of the EE; therefore, the 3D MOT is expected to withstand the launch.

A series of three trigger signals will be sent to the experiment. The first one is to turn off mobile parts of electronic devices such as fans. A second trigger will be sent to start loading the 3D MOT about 1 s before launch. The last trigger will be sent when microgravity is reached to start the experimental sequence in microgravity. 
The experimental sequence in microgravity is illustrated in Fig.~\ref{fig:MicrogravitySetup} c) and detailed in table \ref{tab:microg_seq}. The microgravity sequence starts with the creation of an on-chip BEC using radio-frequency (rf) evaporation. The rf field is deployed by a microscopic U-shaped antenna on the atom chip. This step is estimated to take 1.4 s. It was demonstrated in the MAIUS-1 mission \cite{becker_space-borne_2018,lachmann_ultracold_2021} the creation of a BEC in 1.6 s including 200 ms for the 3D MOT loading. Afterward, the BEC is transported 1 mm away from the atom chip to the center of the interferometry laser beam axis, which will be accomplished via a nonlinear magnetic transport ramp \cite{corgier_fast_2018}.

To reduce the expansion rate of the BEC and benefit from the approximately $2\ \text{s}$ left of microgravity for interferometry, delta kick collimation (DKC) will be used \cite{muntinga_interferometry_2013,chu_proposal_1986,ammann_delta_1997}. The atoms are then transferred to the magnetic insensitive state $m_F=0$ by adiabatic rapid passage using rf fields deployed by the atom chip \cite{dupont-nivet_microwave-stimulated_2015}. The source preparation is estimated to require $1570\ \text{ms}$ over the $4\ \text{s}$ available in microgravity.\\

\begin{figure}[htbp]
\centering
\includegraphics[]{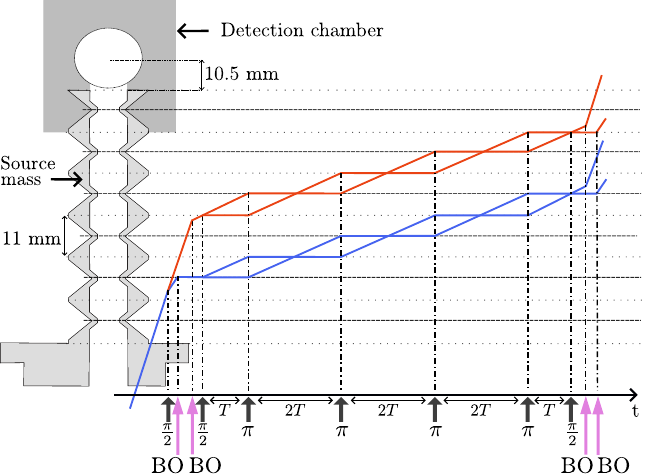}
\caption{Illustration of the implementation of the dual multiloop atom interferometers. The BEC freely propagates until a first beam splitter pulse separates the BEC into two clouds with a momentum difference of $4\hbar k$. The slowest cloud is decelerated via BO to be at rest at the initial position of the first atom interferometer (blue). Meanwhile, the fastest cloud keeps propagating until it is slowed via BO to a final momentum of $4\hbar k$. A beam splitter pulse starts the interferometry sequence once the atomic cloud reaches the initial position of the second interferometer (red). At the end of the interferometry sequence, BO accelerates the output ports so that they reach the detection region before the end of the microgravity time.}
\label{fig:AIscheme}
\end{figure}

Once the atoms are prepared in a non-magnetic and ultra-low expanding state, they need to be transported to the source mass. To this end, Bloch oscillations (BO) will be implemented with the interferometry laser for coherent transport of the BEC \cite{absil_long-range_2023}. 
The atoms are first accelerated with BO for about $1$ ms, at the end of the acceleration phase, the atoms freely propagate to the source mass. The end of the transport is illustrated Fig.~\ref{fig:AIscheme}, a beam splitter pulse separate the BEC in two clouds, one propagating $4v_r$ slower. The slowest cloud is decelerated by BO to be at rest at the initial position of the first interferometer. Meanwhile, the fastest cloud keeps propagating freely until it is decelerated by BO to $4\hbar k$ before reaching the initial position of the second interferometer. Once the cloud reaches the initial position of the second interferometer, a beam splitter pulse starts the dual multiloop atom interferometry. 

At the end of the interferometry sequence, the two output ports at rest and the slow propagation velocity of the two other ports require the implementation of BO acceleration of the output ports such that they can be detected before the end of the microgravity time. This is done by first accelerating the $4\hbar k$ output ports and then accelerating the $0\hbar k$ ouptut ports.
The spatially separated output ports will then be detected by absorption detection \cite{reinaudi_strong_2007,veyron_quantitative_2022}.

\begin{table}[h]
\caption{Typical experimental sequence planned during microgravity.}\label{tab:microg_seq}%
\begin{tabular}{@{}ll@{}}
\toprule
Steps & Time (ms)  \\
\midrule
BEC creation & 1400 \\ 
Magnetic transport & 150  \\ 
Delta kick collimation & 10  \\
Adiabatic rapid passage & 10  \\
Bloch oscillation transport & 200  \\
Multiloop atom interferometry & 1900  \\
Transport/detection & 150  \\
\botrule
Sum & 3820 \\
\end{tabular}
\end{table}

\newpage
\subsection{Vacuum chamber and atom chip}
The periodic source mass and an additional detection chamber have been added to the vacuum system of the MAIUS-A payload see Fig. \ref{fig:MicrogravitySetup} b). An additional non-evaporative getter pump (SAES capacitor with customized modification) has been added to ensure an ultrahigh vacuum in the source mass and detection chamber.

The atom chip has been rotated by 45$\,^{\circ}$ compared to its former orientation in the MAIUS-A setup. Thus, the weak axis of the magnetic trap is orthogonal to the source mass symmetry axis which allows interferometry experiments through the source mass on ground. To reduce external magnetic fields, especially earth magnetic field and stray fields from the Einstein-Elevator, the whole vacuum chamber is enclosed in a three-layer cylindrical magnetic shield \cite{kubelka-lange_three-layer_2016}.

The electronic system and software developed and used for the MAIUS mission \cite{weps2018model} are reused to control and provide the different currents, voltages, and triggers needed for the experimental sequences.
However, the laser system used during the mission MAIUS-1 had to be replaced. The new laser system is described in the following section.

\subsection{Laser system}
In this section, the key subsystems of the DESIRE laser system are presented. The DESIRE laser system is based on the MAIUS-A laser architecture \cite{schkolnik2016compact}, and therefore only a short description will be given here. Some parts of the MAIUS-A laser system, namely the custom assembly fiber splitters and the electronics for beat note detection, have been repurposed in the DESIRE laser system. However, all of the modules mentioned below are new and were produced using commercial components. These modules are: 
\begin{itemize}
    \item The laser sources module, which contains three science lasers for cooling and repumping of the atoms and the master oscillator (MO).
    \item The interferometry laser module which produces the light for Bloch oscillations and Bragg interferometry.
    \item The frequency stabilization module for laser locking.
    \item The switching and distribution module for light distribution, intensity switching, and pulse shaping.
\end{itemize}
The laser sources operate at 780 nm and all optics are mounted on aluminium optical breadboards with walls and covers to protect the system from dust and stray light.
\begin{figure}[htbp]
    \centering
    \includegraphics[width=12 cm]{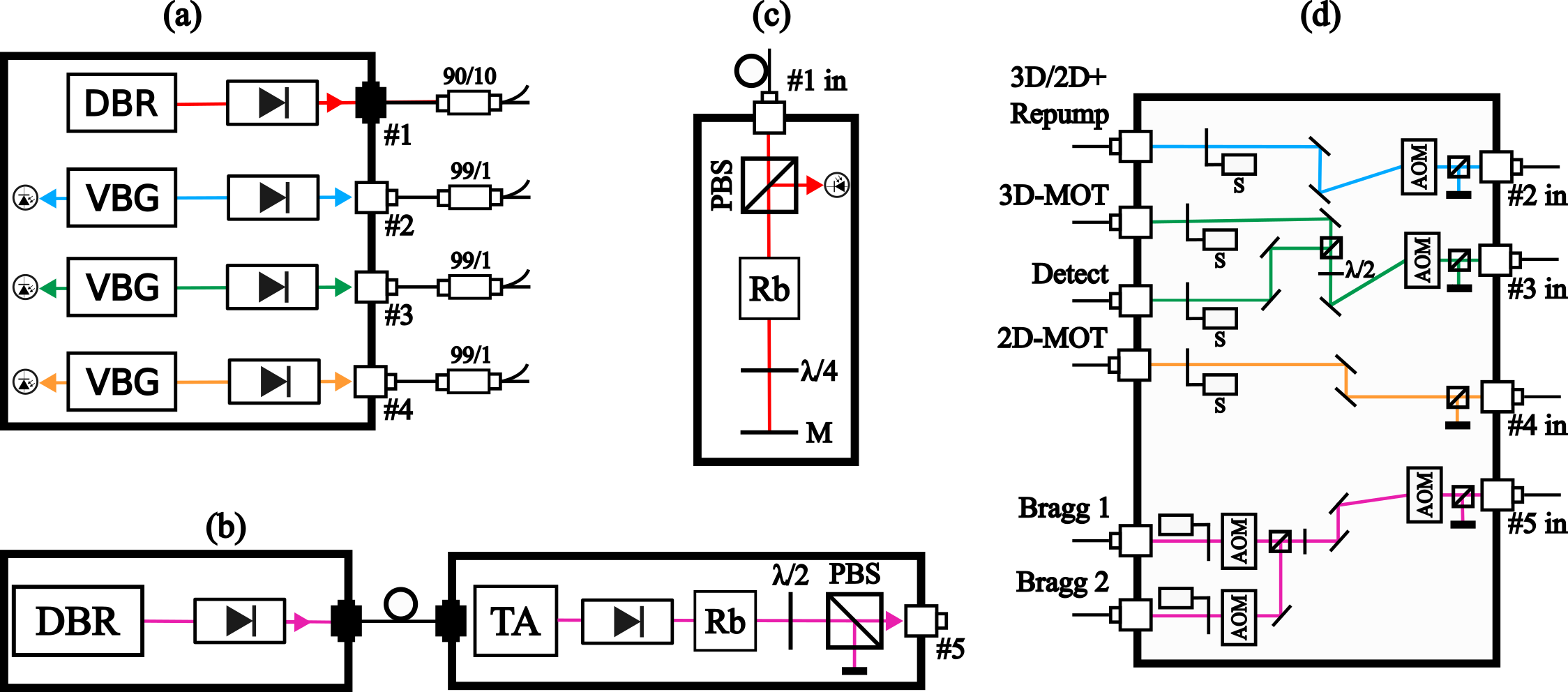}
    \caption{The DESIRE laser system. (a) Laser sources module. (b) Interferometry laser module. (c) Frequency stabilization module. (d) Switching and distribution module.}
    \label{fig:laser system}
\end{figure}

\subsubsection{Laser sources module}
The DESIRE laser module has a footprint of $264\times250$ mm and consists of four 14-pin, type 1 butterfly laser diodes; three free-space Volume Bragg Grating (VBG) stabilised lasers, with a linewidth of about 1.5 MHz and optical power of 500 mW, are employed as science lasers, whereas a fiber-coupled Distributed Bragg Reflector (DBR) laser, with a linewidth of about 1 MHz and optical power of 40 mW, is used as the MO. All of the butterfly laser diodes are placed inside a custom-made aluminium housing with a footprint of $50\times68$ mm. Each housing includes integrated printed circuit boards (PCBs). for current and temperature control. In the case of the three science lasers, there is optical power emitted through their back facet which is monitored with photodiodes. To protect the science lasers from optical feedback, optical isolators are placed directly at their main output. The light from each laser is then coupled directly into PM fibers after passing a half-wavelength wave plate. The fiber system which guides light from the laser module to all other subsystems uses the same fiber splitter system as MAIUS-A in the same configuration \cite{schkolnik2016compact}.

\subsubsection{Interferometry laser module}
The interferometry laser module produces the light for Bloch oscillations and Bragg pulses. For this reason, a Tapered Amplifier (TA) with 3 W output power, seeded by a DBR laser, is employed. To reduce losses due to spontaneous emission, the DBR laser is 100 GHz detuned from the $^{87}$Rb D2 line via a compact Fizeau Wavemeter (FZW). Between the seed laser and the TA, a stage with two optical isolators is employed to ensure the elimination of optical feedback, which could cause mode hops. The free space output of the TA also passes through two isolators to reduce optical feedback in the interferometry laser module. It has been shown that such laser systems suffer from unavoidable excess noise, which exists mainly due to the amplified spontaneous emission (ASE) in the gain medium. To counteract this, a rubidium vapor cell is used to filter out the ASE of the laser around the D2 line of rubidium \cite{andia2015bloch}. The light then is coupled into PM fibers after passing a wave plate and polarizing beam splitter (PBS) to regulate optical power.

\subsubsection{Frequency stabilization module}
At the laser module, the light from each laser is coupled into polarization maintaining fiber splitters with a splitting ratio of 1:99 for the science lasers and 10:90 for the MO laser. The smaller fraction of the light from the MO is guided to a Frequency Modulation Spectroscopy (FMS) module (footprint of $70\times170$ mm). The spectroscopy module is used for stabilizing the MO to the $|F=3\rangle \xrightarrow{} |F'=3/4\rangle$ crossover transition of $^{85}$Rb using Doppler-free FMS. The main part of the light from the MO is split into three parts which are overlapped with the 1\% light from each of the science lasers in an all-fiber-based splitter system. Light from each one of the overlapped outputs is guided onto fast photodiodes adhesively bonded to fiber collimators. The generated beat-notes are used for frequency offset stabilization of the science lasers. 

\subsubsection{Switching and distribution module}
The distribution of the light to the different fibers connected to the vacuum chamber as well as intensity switching and pulse shaping is performed in the switching and distribution module. The module is realized on a $400\times300$ mm optical breadboard. The functionality and pulse shaping of the switching and distribution module are described in \cite{schkolnik2016compact}. There are four input (three science and one interferometry laser) and six output fibers. The science lasers provide the light for repumping, 2D/3D cooling, and detection, whereas the interferometry laser provides the light for the two Bragg beams. The system delivers 107 mW and 25 mW fiber coupled light for cooling and re-pumping, respectively, to the 2D+ MOT and 100 mW and 33 mW for the 3D MOT. The interferometry laser can provide up to 300 mW per fiber.

\section{Conclusion}\label{sec:conclusion}
In this article, we presented the DESIRE project, which aims to test chameleon field theory as a candidate for dark energy by adapting the MAIUS-A payload for use in the Einstein-Elevator facility in Hannover. We outlined the experimental principle, which relies on a spatially structured source mass. The induced gravitational potential and the hypothetical chameleon potential both exhibit spatial periodicity, detectable through multiloop atom interferometry. We explained how the addition of tantalum rings to the source mass modifies the gravitational potential without affecting the chameleon potential, enabling validation of the measurement scheme. We also described how the experimental setup is being adapted for operation in the Einstein-Elevator and detailed typical intended experimental sequences. Finally, we discussed the expected parameter ranges the experiment could exclude, emphasizing its potential to significantly enhance the constraints on chameleon field theory and other light scalar field models, such as the symmetron.

\backmatter
\bmhead{Acknowledgements}
This work was carried out in part at the Jet Propulsion Laboratory, California Institute of Technology, under a contract with the National Aeronautics and Space Administration. This research is also supported by the German Space Agency within the German Aerospace Center (DLR) and the Federal Ministry for Economic Affairs and Climate Action (BMWK) on the basis of a decision by the German Bundestag (FKZ: 50WM2155 and 50WM2156). The authors would also like to thank the DFG and the Lower Saxony state government for their financial support for building the Hannover Institute of Technology (HITec) and the Einstein-Elevator (NI1450004, INST 187/624-1 FUGB) as well as the Institute for Satellite Geodesy and Inertial Sensing of the German Aerospace Center (DLR-SI) for the development and the provision of the experiment carrier system.

\bmhead{Funding}



\section*{Declarations}
\bmhead{Competing interests}
The authors declare no competing interests.
\bmhead{Ethics approval and consent to participate}
Not applicable
\bmhead{Author contribution}
CG, MM, SG, IP, S-wC, AH contributed equally to the writing of the manuscript. All authors read, discussed and approved the final manuscript.
\bmhead{Consent for publication}
Not applicable
\bmhead{Availability of data and materials}
Not applicable. For all requests relating to the paper, please contact author Charles Garcion.
\bmhead{Code availability}
Not applicable

\bigskip


\begin{appendices}






\end{appendices}



\begin{thebibliography}{10}
\providecommand{\doi}[1]{\url{https://doi.org/#1}}
\bibcommenthead

\bibitem[\protect\citeauthoryear{Riess et~al.}{1998}]{riess_observational_1998}
Riess AG, Filippenko AV, Challis P, Clocchiatti A, Diercks A, Garnavich PM,
  et~al.
\newblock Observational evidence from supernovae for an accelerating universe
  and a cosmological constant.
\newblock The astronomical journal. 1998;116(3):1009.

\bibitem[\protect\citeauthoryear{Perlmutter
  et~al.}{1999}]{perlmutter_measurements_1999}
Perlmutter S, Aldering G, Goldhaber G, Knop RA, Nugent P, Castro PG, et~al.
\newblock Measurements of $\Omega$ and $\Lambda$ from 42 high-redshift
  supernovae.
\newblock The Astrophysical Journal. 1999;517(2):565.

\bibitem[\protect\citeauthoryear{Weinberg}{1989}]{weinberg_cosmological_1989}
Weinberg S.
\newblock The cosmological constant problem.
\newblock Reviews of modern physics. 1989;61(1):1.

\bibitem[\protect\citeauthoryear{Martin}{2012}]{martin_everything_2012}
Martin J.
\newblock Everything you always wanted to know about the cosmological constant
  problem (but were afraid to ask).
\newblock Comptes Rendus Physique. 2012;13(6-7):566--665.

\bibitem[\protect\citeauthoryear{Joyce et~al.}{2016}]{joyce_dark_2016}
Joyce A, Lombriser L, Schmidt F.
\newblock Dark energy versus modified gravity.
\newblock Annual Review of Nuclear and Particle Science. 2016;66(1):95--122.

\bibitem[\protect\citeauthoryear{Williams
  et~al.}{2004}]{williams_progress_2004}
Williams JG, Turyshev SG, Boggs DH.
\newblock Progress in lunar laser ranging tests of relativistic gravity.
\newblock Physical Review Letters. 2004;93(26):261101.

\bibitem[\protect\citeauthoryear{Khoury and
  Weltman}{2004}]{khoury_chameleon_2004}
Khoury J, Weltman A.
\newblock Chameleon fields: Awaiting surprises for tests of gravity in space.
\newblock Physical review letters. 2004;93(17):171104.

\bibitem[\protect\citeauthoryear{Adelberger
  et~al.}{2003}]{adelberger_tests_2003}
Adelberger E, Heckel B, Nelson A.
\newblock Tests of the Gravitational Inverse-Square Law.
\newblock Annual Review of Nuclear and Particle Science. 2003;53(1):77--121.

\bibitem[\protect\citeauthoryear{Yin et~al.}{2022}]{yin_experiments_2022}
Yin P, Li R, Yin C, Xu X, Bian X, Xie H, et~al.
\newblock Experiments with levitated force sensor challenge theories of dark
  energy.
\newblock Nature Physics. 2022;18(10):1181--1185.

\bibitem[\protect\citeauthoryear{Burrage et~al.}{2015}]{burrage_probing_2015}
Burrage C, Copeland EJ, Hinds E.
\newblock Probing dark energy with atom interferometry.
\newblock Journal of Cosmology and Astroparticle Physics. 2015;2015(03):042.

\bibitem[\protect\citeauthoryear{Jaffe et~al.}{2017}]{jaffe_testing_2017}
Jaffe M, Haslinger P, Xu V, Hamilton P, Upadhye A, Elder B, et~al.
\newblock Testing sub-gravitational forces on atoms from a miniature in-vacuum
  source mass.
\newblock Nature Physics. 2017;13(10):938--942.

\bibitem[\protect\citeauthoryear{Sabulsky
  et~al.}{2019}]{sabulsky_experiment_2019}
Sabulsky DO, Dutta I, Hinds E, Elder B, Burrage C, Copeland EJ.
\newblock Experiment to detect dark energy forces using atom interferometry.
\newblock Physical Review Letters. 2019;123(6):061102.

\bibitem[\protect\citeauthoryear{Panda et~al.}{2024}]{panda2024measuring}
Panda CD, Tao MJ, Ceja M, Khoury J, Tino GM, M{\"u}ller H.
\newblock Measuring gravitational attraction with a lattice atom
  interferometer.
\newblock Nature. 2024;631(8021):515--520.

\bibitem[\protect\citeauthoryear{Burrage and
  Sakstein}{2018}]{burrage_tests_2018}
Burrage C, Sakstein J.
\newblock Tests of chameleon gravity.
\newblock Living reviews in relativity. 2018;21:1--58.

\bibitem[\protect\citeauthoryear{{{National Institute of Standards and
  Technology (NIST)}}}{2024}]{NISTG}
{{National Institute of Standards and Technology (NIST)}}.: Newtonian constant
  of gravitation.
\newblock Available from \url{https://physics.nist.gov/cgi-bin/cuu/Value?bg}.

\bibitem[\protect\citeauthoryear{Chiow and Yu}{2018}]{chiow_multiloop_2018}
Chiow Sw, Yu N.
\newblock Multiloop atom interferometer measurements of chameleon dark energy
  in microgravity.
\newblock Physical Review D. 2018;97(4):044043.

\bibitem[\protect\citeauthoryear{Lotz
  et~al.}{2017}]{lotz_einstein-elevator_2017}
Lotz C, Frob{\"o}se T, Wanner A, Overmeyer L, Ertmer W.
\newblock Einstein-elevator: A new facility for research from $\mu$ to 5.
\newblock Gravitational and Space Research. 2017;5(2):11--27.

\bibitem[\protect\citeauthoryear{Becker et~al.}{2018}]{becker_space-borne_2018}
Becker D, Lachmann MD, Seidel ST, Ahlers H, Dinkelaker AN, Grosse J, et~al.
\newblock Space-borne Bose--Einstein condensation for precision interferometry.
\newblock Nature. 2018;562(7727):391--395.

\bibitem[\protect\citeauthoryear{Storey and
  Cohen-Tannoudji}{1994}]{storey1994feynman}
Storey P, Cohen-Tannoudji C.
\newblock The Feynman path integral approach to atomic interferometry. A
  tutorial.
\newblock Journal de Physique II. 1994;4(11):1999--2027.

\bibitem[\protect\citeauthoryear{Mcguirk et~al.}{2002}]{mcguirk2002sensitive}
Mcguirk JM, Foster G, Fixler J, Snadden M, Kasevich M.
\newblock Sensitive absolute-gravity gradiometry using atom interferometry.
\newblock Physical Review A. 2002;65(3):033608.

\bibitem[\protect\citeauthoryear{Vovrosh et~al.}{2018}]{vovrosh_additive_2018}
Vovrosh J, Voulazeris G, Petrov PG, Zou J, Gaber Y, Benn L, et~al.
\newblock Additive manufacturing of magnetic shielding and ultra-high vacuum
  flange for cold atom sensors.
\newblock Scientific Reports. 2018;8(1):2023.

\bibitem[\protect\citeauthoryear{Buhmann}{2013}]{buhmann_dispersion_2013}
Buhmann SY.
\newblock Dispersion Forces I: Macroscopic quantum electrodynamics and
  ground-state Casimir, Casimir--Polder and van der Waals forces. vol. 247.
\newblock Springer; 2013.

\bibitem[\protect\citeauthoryear{Buhmann}{2013}]{buhmann_dispersion_2013-1}
Buhmann S.
\newblock Dispersion Forces II: Many-Body Effects, Excited Atoms, Finite
  Temperature and Quantum Friction. vol. 248.
\newblock Springer; 2013.

\bibitem[\protect\citeauthoryear{Holmgren
  et~al.}{2010}]{holmgren_absolute_2010}
Holmgren WF, Revelle MC, Lonij VP, Cronin AD.
\newblock Absolute and ratio measurements of the polarizability of Na, K, and
  Rb with an atom interferometer.
\newblock Physical Review A—Atomic, Molecular, and Optical Physics.
  2010;81(5):053607.

\bibitem[\protect\citeauthoryear{Haslinger
  et~al.}{2018}]{haslinger_attractive_2018}
Haslinger P, Jaffe M, Xu V, Schwartz O, Sonnleitner M, Ritsch-Marte M, et~al.
\newblock Attractive force on atoms due to blackbody radiation.
\newblock Nature physics. 2018;14(3):257--260.

\bibitem[\protect\citeauthoryear{Grosse et~al.}{2016}]{grosse_design_2016}
Grosse J, Seidel ST, Becker D, Lachmann MD, Scharringhausen M, Braxmaier C,
  et~al.
\newblock Design and qualification of an UHV system for operation on sounding
  rockets.
\newblock Journal of Vacuum Science \& Technology A. 2016;34(3).

\bibitem[\protect\citeauthoryear{Steck}{2024}]{SteckRb}
Steck DA.: Rubidium 87 D line data.
\newblock Available online at \url{http://steck.us/alkalidata} (revision 2.3.3,
  28 May 2024).

\bibitem[\protect\citeauthoryear{Lachmann
  et~al.}{2021}]{lachmann_ultracold_2021}
Lachmann MD, Ahlers H, Becker D, Dinkelaker AN, Grosse J, Hellmig O, et~al.
\newblock Ultracold atom interferometry in space.
\newblock Nature communications. 2021;12(1):1317.

\bibitem[\protect\citeauthoryear{Corgier et~al.}{2018}]{corgier_fast_2018}
Corgier R, Amri S, Herr W, Ahlers H, Rudolph J, Gu{\'e}ry-Odelin D, et~al.
\newblock Fast manipulation of Bose--Einstein condensates with an atom chip.
\newblock New Journal of Physics. 2018;20(5):055002.

\bibitem[\protect\citeauthoryear{M{\"u}ntinga
  et~al.}{2013}]{muntinga_interferometry_2013}
M{\"u}ntinga H, Ahlers H, Krutzik M, Wenzlawski A, Arnold S, Becker D, et~al.
\newblock Interferometry with Bose-Einstein condensates in microgravity.
\newblock Physical review letters. 2013;110(9):093602.

\bibitem[\protect\citeauthoryear{Chu et~al.}{1986}]{chu_proposal_1986}
Chu S, Bjorkholm J, Ashkin A, Gordon J, Hollberg L.
\newblock Proposal for optically cooling atoms to temperatures of the order of
  $10^{- 6}$ K.
\newblock Optics letters. 1986;11(2):73--75.

\bibitem[\protect\citeauthoryear{Ammann and
  Christensen}{1997}]{ammann_delta_1997}
Ammann H, Christensen N.
\newblock Delta kick cooling: A new method for cooling atoms.
\newblock Physical review letters. 1997;78(11):2088.

\bibitem[\protect\citeauthoryear{Dupont-Nivet
  et~al.}{2015}]{dupont-nivet_microwave-stimulated_2015}
Dupont-Nivet M, Casiulis M, Laudat T, Westbrook CI, Schwartz S.
\newblock Microwave-stimulated Raman adiabatic passage in a Bose-Einstein
  condensate on an atom chip.
\newblock Physical Review A. 2015;91(5):053420.

\bibitem[\protect\citeauthoryear{Absil et~al.}{2023}]{absil_long-range_2023}
Absil L, Balland Y, Dos~Santos FP.
\newblock Long-range temperature-controlled transport of ultra-cold atoms with
  an accelerated lattice.
\newblock New Journal of Physics. 2023;25(7):073010.

\bibitem[\protect\citeauthoryear{Reinaudi et~al.}{2007}]{reinaudi_strong_2007}
Reinaudi G, Lahaye T, Wang Z, Gu{\'e}ry-Odelin D.
\newblock Strong saturation absorption imaging of dense clouds of ultracold
  atoms.
\newblock Optics letters. 2007;32(21):3143--3145.

\bibitem[\protect\citeauthoryear{Veyron
  et~al.}{2022}]{veyron_quantitative_2022}
Veyron R, Mancois V, Gerent JB, Baclet G, Bouyer P, Bernon S.
\newblock Quantitative absorption imaging: The role of incoherent multiple
  scattering in the saturating regime.
\newblock Physical Review Research. 2022;4(3):033033.

\bibitem[\protect\citeauthoryear{Kubelka-Lange
  et~al.}{2016}]{kubelka-lange_three-layer_2016}
Kubelka-Lange A, Herrmann S, Grosse J, L{\"a}mmerzahl C, Rasel EM, Braxmaier C.
\newblock A three-layer magnetic shielding for the MAIUS-1 mission on a
  sounding rocket.
\newblock Review of scientific Instruments. 2016;87(6).

\bibitem[\protect\citeauthoryear{Weps et~al.}{2018}]{weps2018model}
Weps B, L{\"u}dtke D, Franz T, Maibaum O, Wendrich T, M{\"u}ntinga H, et~al.
\newblock A model-driven software architecture for ultra-cold gas experiments
  in space.
\newblock In: Proceedings of the International Astronautical Congress, IAC;
  2018. .

\bibitem[\protect\citeauthoryear{Schkolnik et~al.}{2016}]{schkolnik2016compact}
Schkolnik V, Hellmig O, Wenzlawski A, Grosse J, Kohfeldt A, D{\"o}ringshoff K,
  et~al.
\newblock A compact and robust diode laser system for atom interferometry on a
  sounding rocket.
\newblock Applied Physics B. 2016;122:1--8.

\bibitem[\protect\citeauthoryear{Andia et~al.}{2015}]{andia2015bloch}
Andia M, Wodey {\'E}, Biraben F, Clad{\'e} P, Guellati-Kh{\'e}lifa S.
\newblock Bloch oscillations in an optical lattice generated by a laser source
  based on a fiber amplifier: decoherence effects due to amplified spontaneous
  emission.
\newblock JOSA B. 2015;32(6):1038--1042.

\end{thebibliography}
\end{document}